\documentclass[%
 aip,
 rsi,
 amsmath,amssymb,
 nofootinbib,
 reprint,%
]{revtex4-2}

\usepackage{graphicx}
\usepackage{dcolumn}
\usepackage{bm}
\usepackage[utf8]{inputenc}
\usepackage[T1]{fontenc}
\usepackage{mathptmx}
\usepackage{etoolbox}
\usepackage{algorithm}
\usepackage{algpseudocode}
\usepackage{booktabs}
\usepackage{hyperref}
\usepackage{cleveref}
\usepackage{xcolor}

\newcommand{\vect}[1]{\bm{#1}}           % bold vector
\newcommand{\mat}[1]{\mathbf{#1}}        % bold matrix
\newcommand{\ray}{\vect{r}}              % ray vector
\newcommand{\pnt}{\vect{p}}              % point vector
\newcommand{\knob}{\vect{\theta}}        % knob-space vector
\newcommand{\func}{\vect{y}}             % beam-space vector
\newcommand{\Jac}{\mat{J}}              % Jacobian
\newcommand{\RTM}{\mat{M}}              % ray transfer matrix
\newcommand{\Trans}{\mat{T}}            % translation matrix
\newcommand{\Rot}{\mat{R}}              % rotation matrix
\newcommand{\comment}[1]{}

\begin{document}

\title{Automatic Optical Alignment Using Projective Geometry}

\author{Bowen Li}
\altaffiliation{Author to whom correspondence should be addressed: bl254@stanford.edu}
\affiliation{Department of Physics, Stanford University, Stanford, California 94305, USA}

\author{Lukas Palm}
\affiliation{Department of Physics, The University of Chicago and the James Franck Institute, Chicago, Illinois 60637, USA}

\author{Xin Wei}
\affiliation{Department of Applied Physics, Stanford University, Stanford, California 94305, USA}

\author{Marius J\"urgensen}
\affiliation{Department of Physics, Stanford University, Stanford, California 94305, USA}

\author{Zeyang Li}
\affiliation{Department of Applied Physics, Stanford University, Stanford, California 94305, USA}

\author{Yiming Cady Feng}
\affiliation{Department of Applied Physics, Stanford University, Stanford, California 94305, USA}

\author{Abhishek V. Karve}
\affiliation{Department of Applied Physics, Stanford University, Stanford, California 94305, USA}

\author{Jon Simon}
\affiliation{Department of Physics, Stanford University, Stanford, California 94305, USA}
\affiliation{Department of Applied Physics, Stanford University, Stanford, California 94305, USA}

\date{\today}

\begin{abstract}
Aligning and maintaining complex optical beam paths is a central challenge across experimental science, because it is a high-dimensional task with strong cross-coupling between controls, often in systems with limited physical access. We present an automated hardware-software framework that resolves this alignment challenge using low-cost, retro-fittable motorized mounts driven by projective-geometry models and a photodiode-fed optimizer. A compact forward model describes the beam path to paraxial order with only the physical mirror angles left free, so it can be rapidly ($\sim$ms) numerically inverted to return the required mirror angles for a desired beam trajectory. A photodiode-fed optimizer then fine-tunes this geometric starting point, and converged mirror settings are tabulated for retrieval in milliseconds and actuation in seconds. We experimentally demonstrate the performance of this approach on a retro-reflected lattice atom-transport system, yielding improvements in both speed and precision over manual alignment. This framework reduces the manual effort required to align complex beam paths, enables programmable optical control in experiments with limited physical access, and enhances the scalability of complex optical architectures.

\end{abstract}

\maketitle

% ============================================================================
%  I. INTRODUCTION
% ============================================================================
\section{\label{sec:intro}Introduction}

% --- The problem (general) --- 
In optics experiments, aligning a laser beam through a system of mirrors, lenses, and apertures is a recurring and time-consuming task. The primary issue is the following: typically only a few elements can be adjusted via knobs (e.g. mirror angles, window plate angles), but those alignment knobs (which we call the ``knob space'') do not directly correspond to a pure position or direction change of the beam at the target plane (which we call the ``beam space''). Even if this mathematical inverse problem is solved, executing it manually---such as turning multiple knobs simultaneously at different speeds for a parallel translation---is beyond human dexterity. As a workaround, researchers rely on manual ``beam walks''. This iterative method is typically limited to two degrees of freedom and demands minute adjustments to avoid losing the feedback signal. The task grows yet more complex when the optical path is housed in inaccessible environments, like vacuum chambers, or when reliant on indirect feedback (e.g., from atoms), which is often slow and information-poor.

% --- Prior work and shortcomings ---
Existing automated approaches each address only part of the difficulty. Gradient-free optimizers (Nelder--Mead~\cite{nelderSimplexMethodFunction1965}, Powell~\cite{powellEfficientMethodFinding1964}) act on a scalar signal: they are slow, blind to the underlying physics, and cannot cold start once the signal is lost. Machine-learning and reinforcement-learning methods~\cite{qinAutomatedAlignmentOptical2025,mathewRaspberryPiAutoaligner2021,uddinAIDrivenRoboticsOptics2025,rakhmatulinReviewAutomationLaser2024} are data-hungry and setup-specific, potentially requiring additional training after configuration changes. General-purpose ray-tracing packages such as Zemax and CODE V provide extensive optical-design capabilities, but their application to closed-loop laboratory alignment requires integration with experimental control hardware and software. Motorized mounts and auto-aligners supply the actuation but not the alignment intelligence. To our knowledge, no lightweight, model-based, real-time alignment system with deterministic model inversion has been demonstrated.

% --- Our approach and how it works ---

Here, we introduce such a deterministic, model-based framework, built on projective geometry~\cite{corcovilosABCDsBetterMatrix2023}, that directly inverts the forward model to return the mirror angles for a desired beam position. Our model operates in stages, each stage having two constraints (e.g. two points through which the beam should go) and two tunable parameters (e.g. alignment knobs of two mirrors). The stages are solved sequentially, and model inversions can be performed in milliseconds. Besides the model's ability to cold start, we demonstrate that a model-free spiral-descent search recovers retro coupling in the tested cases starting far from the optimum. We validate the framework using an eight-degree-of-freedom, retro-reflected transport lattice in an atom--cavity apparatus, using retrofittable, low-cost motorized mounts and a photodiode. The framework enables us to maintain overlap of forward- and backward-transport lattice paths as the atom transport trajectory is displaced over $\sim$1~mm. This automation enables exploration of parameter spaces inaccessible to manual alignment.

% --- Paper organization ---
The paper is organized as follows:
Section~\ref{sec:knobfunc} introduces the knob-space versus beam-space distinction using a single pair of mirrors as the example.
Section~\ref{sec:model} introduces the first-order forward model $g(\knob)$ using $3\times3$ projective geometry, specified directly in lab-frame coordinates and compact enough for real-time evaluation.
Section~\ref{sec:solver} presents a two-stage deterministic solver that splits the four-mirror inversion into two sequential two-dimensional solves.
Section~\ref{sec:correction} introduces additional corrections to account for deviations from the model: a photodiode-fed optimizer refines the model solution and builds a correction table that later alignments recall by interpolation.
Section~\ref{sec:experiment} reports the experimental validation on the eight-degree-of-freedom alignment of the transport system in an atom--cavity apparatus.
All code, CAD files, and firmware are released as open source in Ref.~\onlinecite{motorizedmirror}.

\section{\label{sec:knobfunc}Knob vs. beam space}

\begin{figure}
  \includegraphics[width=\columnwidth]{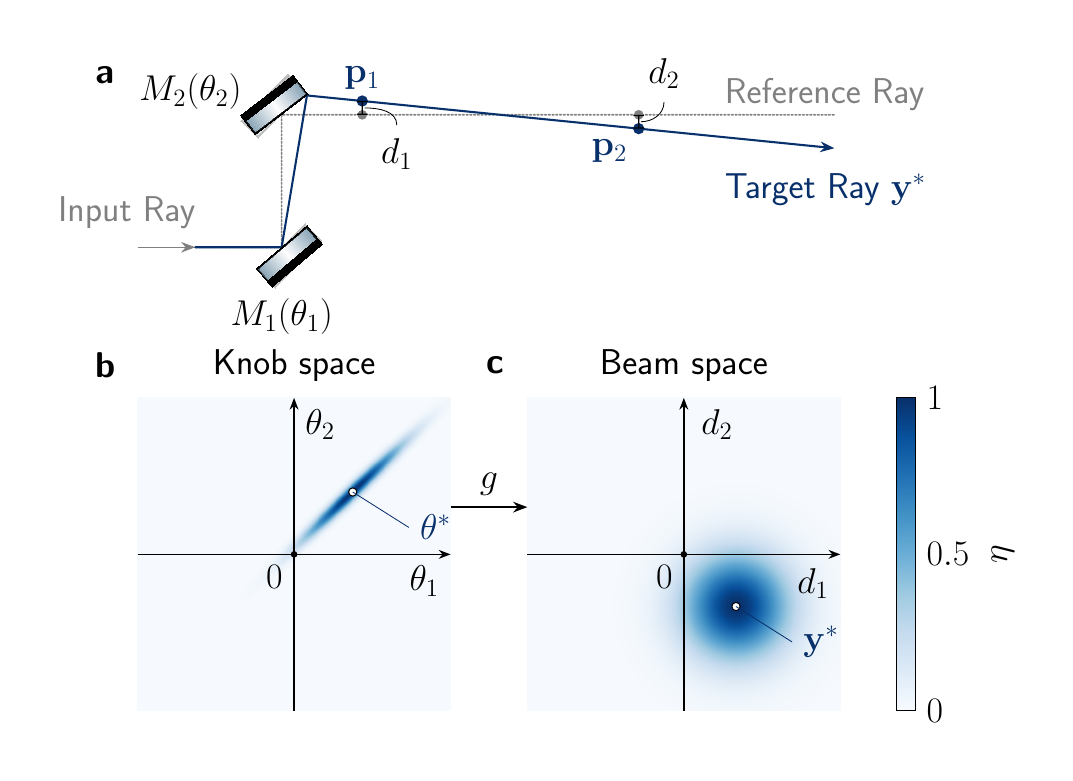}
  \caption{\label{fig:ellipse}
    \textbf{Demonstration of a two-mirror pair beam walk}. 
    (a) Mirrors $M_1,M_2$ steer the input beam through two target points, $\pnt_1,\pnt_2$, which sit at predetermined axial positions and are specified by their transverse offsets $d_1,d_2$ from the nominal beam axis (gray line). A desired target ray, $\func^*=(d_1^*,d_2^*)$ (blue), is realized by the solved mirror angles $\theta_1,\theta_2$.
    (b),(c) Beam coupling in two coordinate systems. The heatmaps plot the deviation from the target ray, defined as $\eta = \exp\!\left[-(d_1-d_1^*)^2-(d_2-d_2^*)^2\right]$. We use a Gaussian exponential to simulate mode overlap, as it would occur, e.g., when the steered beam is coupled into an optical fiber. Perfect alignment gives $\eta = 1$.
    (b)~In \emph{knob space}, $\knob=(\theta_1,\theta_2)$, the coupling forms a narrow ellipse whose aspect ratio is the Jacobian condition number $\kappa \gg 1$, making manual optimization difficult and blind to the correlation structure.
    (c)~In \emph{beam space}, $\func=(d_1,d_2)$, the parameters are decoupled and the coupling forms an isotropic Gaussian ($\kappa\approx1$).
  }
\end{figure}

We illustrate the essential challenge of alignment using a single two-mirror pair in Fig.~\ref{fig:ellipse}(a). The two mirrors, $M_1$ and $M_2$, are employed to steer a beam through two target points $\pnt_1$ and $\pnt_2$, which lie in two control planes at predetermined axial positions along the reference axis, so that only their transverse offsets are free. Moving these target points transversely---that is, changing the beam's position and direction---demands a coordinated ``walk'' of both mirrors' knob angles with small enough knob adjustments to ensure the optimization signal is not lost. The knobs of the mirrors form the ``knob space'' $\knob $, while the beam's transverse offsets $(d_1, d_2)$ from the reference axis at the control planes form the ``beam space'' $\func $. The forward map $\func = g(\knob)$ takes the mirror knob angles to these offsets, and alignment is its inversion,
\begin{equation}\label{eq:inversion}
  \knob^* = g^{-1}(\func^*),
\end{equation}
where $\knob^*$ are the knob angles that lead to the desired target beam $\func^*$. Formally, the Jacobian $\Jac = \partial \func / \partial \knob$ generically has a large ratio of its largest to smallest singular value, i.e., a large condition number $\kappa$ (derived for a two-mirror pair in Appendix~\ref{app:jacobian}), meaning that input errors are amplified under inversion. Intuitively, this can also be inferred from Fig.~\ref{fig:ellipse}(a), as the mirror knobs have different sensitivities based on their distance from the target. When evaluating the deviation from the target ray as a function of the mirror knobs, this leads to a strongly elongated response ellipse [Fig.~\ref{fig:ellipse}(b)], necessitating a tedious beam walk. Such ill-conditioning is generic: assigning the beam position to one mirror knob and the beam angle to another requires the optical system to be designed around the control mirrors, e.g., with Fourier-transform lenses placed between them. In practice, both mirrors instead affect beam position and beam angle to a similar degree, such that a pure change of position or angle demands large, coordinated adjustments of both mirrors. In contrast, in beam space, the response is symmetric as shown in Fig.~\ref{fig:ellipse}(c). The problem only compounds when more mirrors and target points are chained together: the degrees of freedom multiply and their couplings grow, making the landscape both higher-dimensional and more strongly coupled.

% ── Figure 1: Ellipse → Circle ──

\begin{figure*}
  \includegraphics[width=\textwidth]{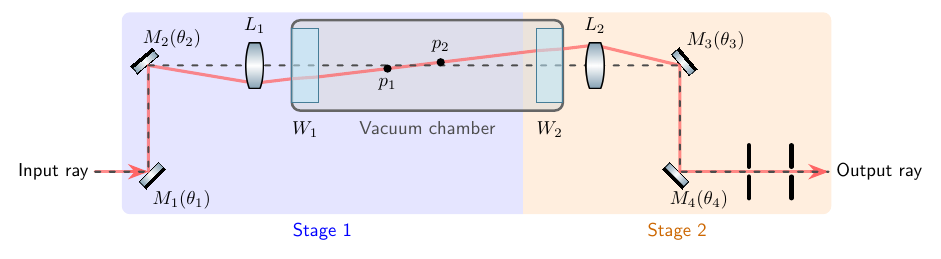}
  \caption{\label{fig:layout}%
    \textbf{Optical layout of the four-mirror alignment setup in one plane.} After exiting the input fiber, the first two-mirror pair (mirrors $M_1$, $M_2$) steers the beam (red) through a telescope lens $L_1$ into the vacuum chamber (with entrance window $W_1$). In the vacuum chamber, the beam passes through the two target points $\pnt_1$ and $\pnt_2$. It then exits through the exit window $W_2$ and lens $L_2$, before it is steered by a second two-mirror pair ($M_3$, $M_4$) through a pair of constrained apertures such that the output ray stays fixed.
    The background shading indicates the two solver stages: Stage~1 aligns the beam to $\pnt_1$, $\pnt_2$; Stage~2 holds the output ray fixed on the
    design reference axis (black dashed line). All knobs are actuated through servo or piezo motors (see Appendix~\ref{app:hardware}).
  }
\end{figure*}

Just such a complicated regime motivated the present work: a four-mirror setup (Fig.~\ref{fig:layout}), arranged as two two-mirror pairs---eight degrees of freedom in total ($4\times2$ axes)---must simultaneously thread the beam through two target points, $\pnt_1$ and $\pnt_2$, while keeping the output ray fixed (conceptually indicated by two apertures), all while several further elements, like lenses and the vacuum chamber, are fixed in the beam path. In general, the output ray must be held invariant whenever downstream elements depend upon it; in our apparatus, the beam is retro-reflected to form a transport lattice, and thus must pass through two double-passed acousto-optic modulators (AOMs) at the exact Bragg angle such that the back-reflected beam ultimately couples back into the input fiber.

% ── Figure 2: Optical layout ──

% ============================================================================
%  II. PROJECTIVE GEOMETRY FOR OPTICS
% ============================================================================
\section{\label{sec:model}Projective Geometry for Optics}

Standard $2\times2$ ABCD ray transfer matrices \cite{siegmanLasers1986} describe a beam to first order about a single straight reference axis: every optical element must be centered on that axis and joined to its neighbors by explicit free-space propagation matrices. Misaligned elements can be included by augmenting the matrix with a third column~\cite{siegmanLasers1986}, but the augmented matrices act affinely on the ray's intercept and slope, so a tilted element enters only as a constant slope offset, the first-order stand-in for a rotation. Real setups use folded paths, mirrors at arbitrary angles and positions, periscopes, and off-axis routing through vacuum chambers, whose finite-angle turns act nonlinearly on the slope and so have no augmented-matrix representation; the path must then be unfolded by hand, with all path lengths and per-reflection coordinate flips tracked explicitly.

Instead, we adopt the $3\times3$ homogeneous-coordinate formalism~\cite{corcovilosABCDsBetterMatrix2023}, in which reflections, rotations, and translations are exact, arbitrary lab-frame element placement requires no free-space propagation matrices, and only focusing and refracting elements remain paraxial. In this formalism, a ray traveling along the line $ax + by + c = 0$ in a transverse plane is represented by the homogeneous vector
\begin{equation}\label{eq:ray}
  \ray = \begin{pmatrix} c \\ a \\ b \end{pmatrix}.
\end{equation}
The slope $m = -a/b$ and intercept $h = -c/b$ recover the usual ABCD state
vector $(h, m)^T$. Any optical element has a canonical ray transfer matrix $\RTM_0$. For a mirror, $\RTM_0 = \mathrm{diag}(-1,\,1,\,-1)$; for a thin lens of focal
length~$f$,
\begin{equation}\label{eq:lens}
  \RTM_0 = \begin{pmatrix}
    1 & 0 & 0 \\ -1/f & 1 & 0 \\ 0 & 0 & 1
  \end{pmatrix}.
\end{equation}
Additionally, the lab-frame matrix describes optical elements placed at position $(u,v)$ with orientation $\varphi$, measured counter-clockwise from the canonical orientation:
\begin{equation}\label{eq:sandwich}
  \RTM = \Trans(u,v)\;\Rot(\varphi)\;\RTM_0\;\Rot^{-1}(\varphi)\;\Trans^{-1}(u,v),
\end{equation}
where $\Trans(u,v)$ translates a ray by $(u,v)$ and $\Rot(\varphi)$ rotates it counter-clockwise about the origin by $\varphi$, following Ref.~\onlinecite{corcovilosABCDsBetterMatrix2023} (explicit forms in Appendix~\ref{app:matrices}). For convenience, we define the mirror angles $\theta_i$ as deviations from nominal orientations $\varphi_0$, so that they enter as the orientation $\varphi = \varphi_0 + \theta_i$ in the sandwich transform [Eq.~\eqref{eq:sandwich}] and are the only free parameters in the model.

The formalism also inherits the projective duality of homogeneous coordinates:
\begin{align}
  \text{Line through two points:} \quad
    & \ray = \pnt_1 \times \pnt_2, \label{eq:line_thru_pts} \\
  \text{Intersection of two lines:} \quad
    & \pnt = \ray_1 \times \ray_2. \label{eq:intersection}
\end{align}
Here, the same cross-product operation has dual geometric interpretations.
Crucially, element positions enter through $\Trans(u,v)$ and ray intersections are resolved automatically via Eq.~\eqref{eq:intersection} with no explicit propagation distances needed.

In a three-dimensional optical beam path, the sagittal and tangential planes (associated with the horizontal $x$ and vertical $y$ beam coordinates, respectively) are treated as two independent $3\times3$ systems. The details of the treatment of the tangential plane are described in Appendix~\ref{app:tangent}. For simplicity, all equations below are written in the sagittal plane. For out-of-plane reflections we combine the sagittal and tangential planes in the solver.
This decoupling is well approximated for setups with parallel or $90^\circ$ periscopes and axis-aligned astigmatic elements, which is sufficient for most optical setups. The fully coupled case (arbitrary image rotation) requires a Plücker-coordinate extension~\cite{arnaudGaussianLightBeams1969, nilssonEigenpolarizationTheoryMonolithic1989, pluckerNewGeometrySpace1865, pottmannComputationalLineGeometry2001}.

\section{\label{sec:solver}The Two-Stage Deterministic Solver}

We now return to the alignment problem of Fig.~\ref{fig:layout}. The key structural observation is that each set of constraints is immediately preceded by its own steerable mirror pair: the first mirror pair ($M_1$, $M_2$) steers the beam onto the target points $\pnt_1$ and $\pnt_2$, which fixes the beam's position and direction inside the chamber; the second mirror pair ($M_3$, $M_4$) then steers the outgoing beam through the apertures. These two tasks are the two stages of our solver. Placing a two-mirror pair [Fig.~\ref{fig:ellipse}(a)] before each set of targets is standard practice in optical layouts because it decouples the individual alignment tasks: each pair supplies exactly the number of degrees of freedom needed to satisfy the constraints that follow it. As a consequence, the four-degree-of-freedom inversion $\knob^* = g^{-1}(\func^*)$ [Eq.~\eqref{eq:inversion}] decomposes into two two-dimensional problems---one per stage---solved sequentially in a single forward pass.

Setting up the model is then straightforward: ray propagation along any segment of the beam path is the ordered product of the $3\times3$ ray-transfer matrices of the elements in that segment. Only the four mirror angles $\theta_1$ to $\theta_4$ are kept symbolic; all other elements---the telescope lenses $L_1$, $L_2$ and the vacuum-chamber windows---are fixed and their ray-transfer matrices are constant. Each stage thus reduces to two symbolic mirror matrices sandwiched between constants, a structure that is easily inverted. The solved mirror angles are then sent to the mirror-knob actuators (see Appendix~\ref{app:hardware}). The following sections use the configuration of Fig.~\ref{fig:layout} as a concrete example.

% --- Stage 1 ---
\subsection{\label{sec:stage1}Stage 1: First two-mirror pair---beam through \texorpdfstring{$\pnt_1$ and $\pnt_2$}{p1 and p2}}

The target ray connecting the two target points $\pnt_1$ and $\pnt_2$ is constructed via Eq.~\eqref{eq:line_thru_pts}:
\begin{equation}\label{eq:rtarget}
  \ray_\mathrm{target} = \pnt_1 \times \pnt_2.
\end{equation}
The input ray is propagated symbolically through the first mirror pair $M_1$, $M_2$, followed by the fixed downstream elements (telescope lens $L_1$ and the chamber entrance window $W_1$). Collecting the elements of this segment into a single ray transfer matrix,
\begin{equation}\label{eq:RTMA}
  \RTM_A(\theta_1, \theta_2)
    \equiv \RTM_{W_1}\,\RTM_{L_1}\,\RTM_{M_2}(\theta_2)\,\RTM_{M_1}(\theta_1),
\end{equation}
the propagated ray is
\begin{equation}\label{eq:rmid}
  \ray_\mathrm{mid}(\theta_1, \theta_2)
    = \RTM_A(\theta_1, \theta_2)\;\ray_\mathrm{in}.
\end{equation}
The alignment condition (the propagated ray must coincide with the target ray) is expressed as a collinearity constraint in homogeneous coordinates:
\begin{equation}\label{eq:constraint1}
  \ray_\mathrm{mid}(\theta_1, \theta_2)
    \times \ray_\mathrm{target} = \vect{0}.
\end{equation}
Since the cross product of two three-vectors yields a three-vector with one redundant component (by $\vect{v}\cdot(\vect{v}\times\vect{w}) = 0$), this provides two independent scalar equations in two unknowns $(\theta_1, \theta_2)$. Equation~\eqref{eq:constraint1} is the mechanism by which the model inversion $g^{-1}$ is evaluated.

\subsection{\label{sec:stage2}Stage 2: Second two-mirror pair and output-ray invariance}

Stage~2 has the complementary task: mirror pair $M_3$, $M_4$ re-steers the beam coming from Stage 1 such that the output ray stays co-aligned with the reference path going through the two apertures. Note that such a constraint is imposed by whatever the output ray must still feed downstream; in our apparatus these are the double-pass AOMs and the retro-reflector. The double-pass AOM is both a sensitive angle filter (Bragg condition) and a spatial filter (small aperture for correct diffraction order), and its drive frequency is reserved exclusively for conveyor-belt atom transport~\cite{schraderOpticalConveyorBelt2001} and is not available for alignment correction; the retro-reflector must then couple the beam back into the input fiber. Any deviation of the output ray from the design path causes loss of diffraction efficiency and beam walk-off, destroying the retro-reflected lattice beam. The output ray must therefore be \textit{invariant} with respect to the design-path reference ray.

Analogous to Stage~1, the output ray is given by:
\begin{equation}\label{eq:rout}
  \ray_\mathrm{out}(\theta_3, \theta_4)
    = \RTM_B(\theta_3, \theta_4)\;
      \ray_\mathrm{mid}(\theta_1^*, \theta_2^*),
\end{equation}
where
\begin{equation}\label{eq:RTMB}
  \RTM_B(\theta_3, \theta_4)
    \equiv \RTM_{M_4}(\theta_4)\,\RTM_{M_3}(\theta_3)\,\RTM_{L_2}\,\RTM_{W_2}.
\end{equation}
Let $\ray_\mathrm{ref}$ denote the reference ray, the output ray required by the downstream optics (in our apparatus, the design path through the double-pass AOMs and retro-reflector). The output ray invariance constraint is
\begin{equation}\label{eq:constraint2}
  \ray_\mathrm{out}(\theta_3, \theta_4)
    \times \ray_\mathrm{ref} = \vect{0},
\end{equation}
which has the same structure as Eq.~\eqref{eq:constraint1}: two equations and two
unknowns. The two stages are decoupled and solved sequentially, with no iteration between them.

% --- Solution methods ---
\subsection{Algorithm}

The final algorithm is shown as pseudocode in Algorithm~\ref{alg:solver}. We solve the constraints, Eqs.~\eqref{eq:constraint1} and \eqref{eq:constraint2}, numerically via a bounded least-squares (\texttt{scipy.optimize.least\_squares}, trust-region-reflective algorithm~\cite{2020SciPy-NMeth}) with bounds $|\theta_i| \leq \pi/8$ and tolerances $\mathrm{ftol} = \mathrm{xtol} = \mathrm{gtol} = 10^{-12}$. Convergence is reached in 3--5 iterations (${\sim}$ms on a Raspberry~Pi~5).

For non-gimbal kinematic mirror mounts, the sagittal and tangential paths are coupled: a tilt in one plane slightly displaces the effective reflection point seen by the orthogonal plane. In practice this can often be neglected. If not, a cross-coupling compensation can be used: the $y$-path is solved first, and the resulting mirror offsets are fed as compensation into the $x$-path solver. Because the coupling is reciprocal, the fully self-consistent solution iterates this pass to convergence; a single pass suffices at small tilt angles, while large angles require a few iterations (Appendix~\ref{app:tangent}).

\begin{algorithm}[H]
\caption{Model inversion: $\knob^* = g^{-1}(\func^*)$}
\label{alg:solver}
\begin{algorithmic}[1]
  \Require Element matrices $\{\RTM_i(\theta_i)\}$, input ray $\ray_\mathrm{in}$,
           target points $\pnt_1$, $\pnt_2$,
           reference ray $\ray_\mathrm{ref}$ (fixed by downstream optics)
  \Ensure Mirror angles $\knob^* = (\theta_1^*,\theta_2^*,\theta_3^*,\theta_4^*)$
  \Statex
  \Statex \textbf{--- Stage 1 ---}
  \State $\ray_\mathrm{target} \gets \pnt_1 \times \pnt_2$
  \State $\ray_\mathrm{mid}(\theta_1,\theta_2)
         \gets \RTM_A(\theta_1,\theta_2)\cdot\ray_\mathrm{in}$
  \State Solve $\ray_\mathrm{mid} \times \ray_\mathrm{target} = \vect{0}$
         for $(\theta_1^*,\theta_2^*)$
  \Statex
  \Statex \textbf{--- Stage 2 ---}
  \State $\ray_\mathrm{out}(\theta_3,\theta_4)
         \gets \RTM_B(\theta_3,\theta_4)\cdot\ray_\mathrm{mid}(\theta_1^*,\theta_2^*)$
  \State Solve $\ray_\mathrm{out} \times \ray_\mathrm{ref} = \vect{0}$
         for $(\theta_3^*,\theta_4^*)$
  \Statex
  \Statex \textbf{--- Command ---}
  \State Send $\knob^*$ to motor controller; await encoder confirmation
\end{algorithmic}
\end{algorithm}

% ============================================================================
%  IV. BRIDGING MODEL AND REALITY
% ============================================================================
\section{\label{sec:correction}Bridging Model and Reality}
While the first-order projective model provides a deterministic starting point $\knob_\mathrm{model} = g^{-1}(\func^*)$, it is only paraxial and leaves a residual from mechanical imperfections (mount hysteresis, thermal drift) and unmodeled elements (window wedge, optical aberrations, coordinate twist). Additionally, the modeled element parameters (distances, focal lengths, etc.) can deviate from the experimental values. In practice, the model places the beam close enough to the optimum that the residual landscape is unimodal and correction is fast.

We address this in the following way: First, a one-time \emph{Model calibration} step tunes a few model parameters so that they absorb, to first order, the mismatch between model and apparatus. Whatever residual remains is then handled by the correction stage, which closes the loop against a measured signal rather than the model: an \emph{Optimization} step that drives the beam to the optimum using a live photodiode reading, and an \emph{Interpolation} step that recalls pre-optimized solutions without further measurement. All corrections use the retro-coupled fiber power $P(\knob)$ as a scalar alignment metric, which is measured through a photodiode (Thorlabs FDS100) and an MCP3424 analog-to-digital converter (ADC; 12-bit, 10-sample average) on the Raspberry~Pi~5.

% --- Model calibration ---
\subsection{\label{sec:modelcal}Model calibration}
The model's few physical parameters, such as element positions and focal lengths, are initially set from mechanical measurements. Rather than measuring them individually to higher precision, we treat a small number of them as \emph{effective} parameters: quantities whose fitted values need not reproduce the corresponding physical component, but which are chosen so that a shift in the model compensates, to first order, the imperfections the model omits. The calibration therefore scans a candidate parameter and asks which of its values keeps the model's predictions valid over the largest range of targets. At each value of the parameter, we rebuild the model, re-solve the alignment for a series of displaced target points, and record the retro-coupled fiber power. The value that cancels the first-order mismatch keeps the power high over the widest displacement range.

\begin{figure}
  \includegraphics[width=\columnwidth]{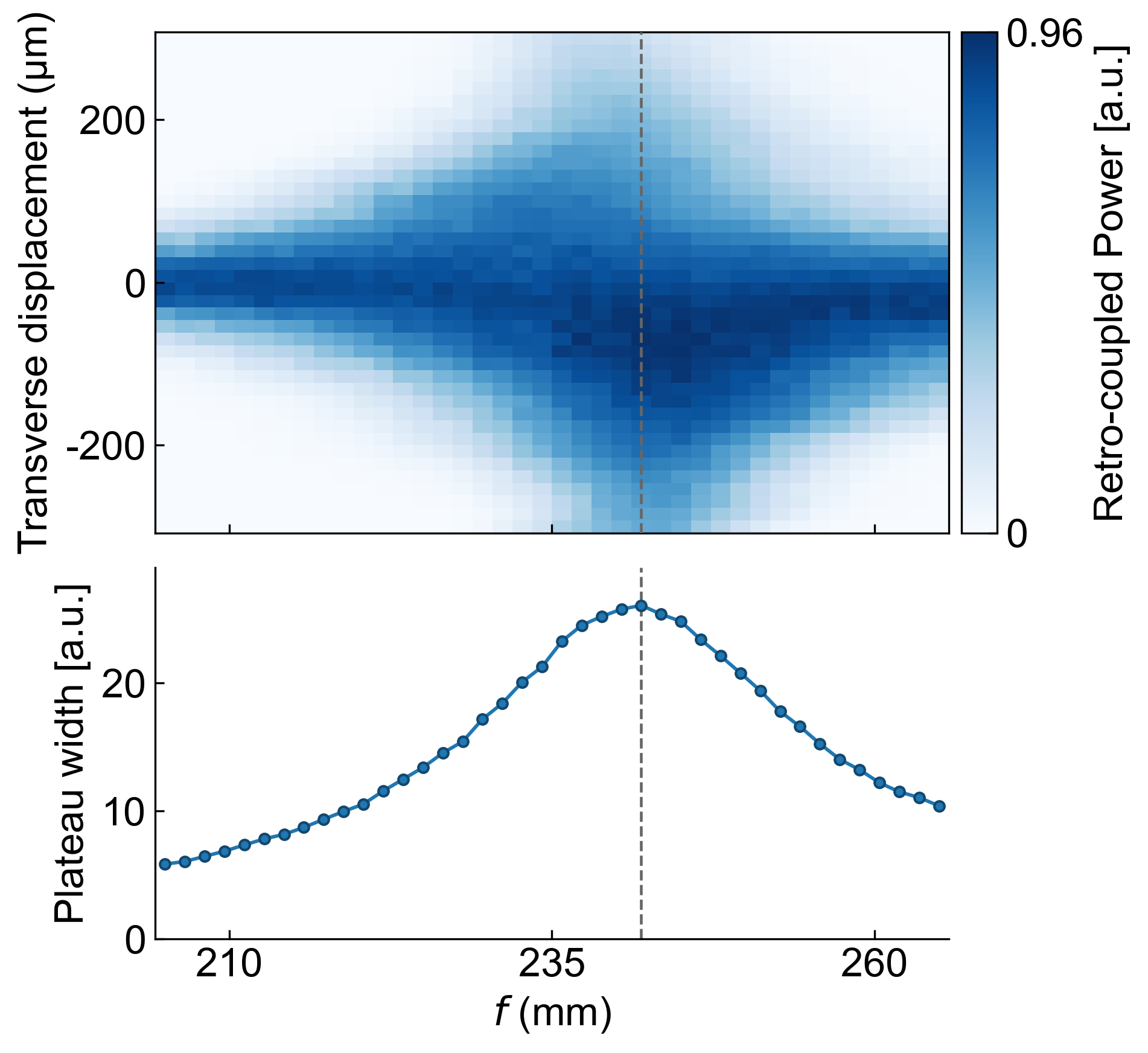}
  \caption{\label{fig:calibration}%
    \textbf{Example of model calibration}.
    Top: retro-coupled fiber power as a function of the transverse displacement of $\pnt_1$ at each modeled focal length, $f$, of telescope lens $L_2$. At each value of modeled focal length the alignment is re-solved as the target point (atom) is displaced, so the power measures the model's ability to keep the retro-fiber alignment (Stage~2) while moving the control point (Stage~1). The plateau is widest for the parameter value that best compensates the model's residual error to first order, and that value is what the calibration returns.
    Bottom: the plateau width, taken as the retro power summed over the displacement scan at each focal length. Its maximum (dashed line) sets the calibrated effective focal length, $f = 241.9$~mm.
    }
\end{figure}

Figure~\ref{fig:calibration} shows this for the focal length of telescope lens $L_2$. We choose the figure of merit as the retro power summed over the displacement scan, which locates the optimum at $f = 241.9$~mm. This value is the one that best linearizes the model around our operating point; any residual, higher-order mismatch is left to the correction stage of Sec.~\ref{subsec:optimize}. The procedure is performed once, scanning one parameter at a time. How sharply the merit function peaks as a parameter is scanned [Fig.~\ref{fig:calibration}, bottom] sets that parameter's tolerance: for most parameters the peak is broad, so they can keep their nominal, mechanically measured values, and only the few with a sharp peak need to be calibrated. In our system this is essentially the telescope focal length. Degeneracies between parameters are benign, since the goal is a model that predicts well over the working range rather than a unique parameter set.

% ── Figure: Model calibration ──

\subsection{Optimization} \label{subsec:optimize}
The Optimize step refines the model guess $\knob_\mathrm{model} = g^{-1}(\func^*)$---or the interpolated guess of the Interpolate step (see the following section)---to the retro-power optimum using live photodiode feedback. It serves two purposes: it populates the Interpolate correction table node by node, and it can be re-run on demand when slow drifts have degraded the stored solution. We use two optimization strategies of increasing generality and cost.

\begin{figure*}
\columnwidth=\textwidth % the algorithm float boxes itself at \columnwidth
\begin{algorithm}[H]
\caption{Spiral-descent search for the retro-power optimum.}
\label{alg:spiral}
\begin{algorithmic}[1]
  \Require measured retro power $P(\vec{\theta})$ over a 2D knob pair
           $\vec{\theta}=(\theta_1,\theta_2)\in\mathbb{R}^2$ (the two mirror knobs of the pair);
           start $\vec{\theta}_0$; spiral spacing $d$; samples per turn $N$;
           drift rate $\alpha\in(0,1]$
  \Ensure best knob setting $\vec{\theta}^\star$ found
  \State $\vec{c}\gets\vec{\theta}_0$;\quad $\vec{\theta}^\star\gets\vec{\theta}_0$;\quad
         $\varphi\gets0$ \Comment{$\vec{c}$: current spiral center}
  \While{evaluation budget remains}
    \Statex \hspace{\algorithmicindent}\emph{(A) step one point outward along the spiral}
    \State $\varphi \gets \varphi + 2\pi/N$;\quad $r \gets d\,\varphi$
           \Comment{radius grows with angle}
    \State $\vec{\theta} \gets \vec{c} + r\,(\cos\varphi,\,\sin\varphi)$, clipped to bounds
           \Comment{Cartesian point at polar radius $r$, angle $\varphi$ about $\vec{c}$}
    \State $P \gets P(\vec{\theta})$ \Comment{move servos, read photodiode}
    \State append $(\vec{\theta},P)$ to the history; keep the last $N$ samples
    \If{$P > P(\vec{\theta}^\star)$}
      \State $\vec{\theta}^\star\gets\vec{\theta}$;\quad $\vec{c}\gets\vec{\theta}$;\quad $\varphi\gets0$
             \Comment{new best: recenter, restart spiral}
    \EndIf
    \Statex \hspace{\algorithmicindent}\emph{(B) slide the center toward higher power}
    \State $\bar{\vec{\theta}} \gets
            \big(\textstyle\sum_k P_k\,\vec{\theta}_k\big)/\big(\sum_k P_k\big)$
           over the last $N$ samples \Comment{power-weighted average}
    \State $\vec{c} \gets \vec{c} + \alpha\,(\bar{\vec{\theta}}-\vec{c})$
           \Comment{move a fraction $\alpha$ toward it}
  \EndWhile
  \State \Return $\vec{\theta}^\star$
\end{algorithmic}
\end{algorithm}
\end{figure*}

The default strategy is a model-free search, robust to the first-order decorrelation degrading as the displacement grows. A Nelder--Mead~\cite{nelderSimplexMethodFunction1965} simplex maximizes $P(\knob)$ on the lower two-mirror pair ($M_3$, $M_4$) only, operating not on the raw knobs but in the pair's \emph{ridge eigen-directions}---the principal axes, \emph{along} the flat ridge and \emph{across} its steep wall, that decorrelate the mirror-pair response just as the model's beam space does [Fig.~\ref{fig:ellipse}(c)]. These directions are cold-start derived in either of two ways: from the model itself, which predicts the ridge orientation, or from a quick power measurement over the two knobs (Appendix~\ref{app:spiral}). We optimize the two knobs in the tangential plane first, then the two knobs in the sagittal plane; each evaluation reads the retro-coupled fiber power, and convergence requires ${\leq}\,12$ evaluations per axis, 15--20~s for the two axes.

When the requested target lies beyond the calibrated grid, it is possible that the Nelder--Mead algorithm stalls. In those cases, we first run a spiral-descent search~\cite{servo_aligner} directly in the raw knob coordinates to acquire and optimize the signal. The spiral search [Algorithm~\ref{alg:spiral}] samples along an Archimedean spiral whose center is dragged toward the intensity-weighted centroid of recent samples, moving towards the optimum while keeping servo moves short. The spiral search can reach the retro optimum when the model prediction yields insufficient retro signal for local refinement. The model can then be re-calibrated around the optimized solution. Moreover, the spiral search can also be used to find the principal direction of the ridge, which can then be fed to the Nelder--Mead optimizer for fast and accurate convergence (Appendix~\ref{app:spiral}). Spiral search followed by Nelder--Mead refinement completes in ${\sim}2$--$4$~min for the two axes in our setup.

% --- Interpolate ---
\subsection{Interpolation}
Finally, we make use of an Interpolate step. In this step, we run the Optimize step once on a four-dimensional grid $\tilde{\func}$ of target positions and store the resulting optimized knob angles $\tilde{\knob}$. At runtime, we then only need to interpolate this table---no model inversion and no optimization is needed. We use a multilinear interpolant over the grid (for example, 6 points per axis, $6^4 = 1296$ nodes, 8 servo angles per node, ${\sim}6$~h) and extrapolation is used beyond the grid bounds. As a single retrieval costs only ${\sim}$ms, repeated alignments require no further solving; the user can optionally re-trigger the Optimize step (from the interpolated start) to refresh a node or the whole table when slow drifts are suspected.

% --- Pipeline summary ---
In summary, the pipeline runs in three modes, all seeded by the same model inversion $\knob_\mathrm{model} = g^{-1}(\func^*)$: a one-time calibration (model inversion plus the Nelder--Mead step, ${\sim}6$~h) that runs the Optimize step at every grid node to build the table; the default runtime recall (the Interpolate step, ${\sim}$ms), which serves any target by interpolation alone; and an optional on-demand Optimize pass (15--20~s) that refreshes a node from the interpolated start when slow drift is suspected. Until a table has been built, the model inversion alone supplies the starting knob angles, allowing alignment from a cold start. The full software/hardware stack implementing this pipeline is summarized in Appendix~\ref{app:architecture}.

\begin{table}[b]
\caption{\label{tab:timing}%
Alignment timing breakdown. All times measured on a Raspberry~Pi~5.}
\begin{ruledtabular}
\begin{tabular}{lc}
  \textrm{Step} & \multicolumn{1}{c}{\textrm{Time}} \\
  \colrule
  Model inversion $g^{-1}(\func^*)$        & \multicolumn{1}{c}{1--5~ms} \\
  Mirror motion (servo)                     & \multicolumn{1}{c}{1--2~s} \\
  Correction---Nelder--Mead (two axes)      & \multicolumn{1}{c}{15--20~s} \\
  Correction---spiral, cold start (two axes) & \multicolumn{1}{c}{${\sim}2$--$4$~min} \\
  \colrule
  \textbf{Total (model $+$ Nelder--Mead)}   & \multicolumn{1}{c}{\textbf{${\sim}20$~s}} \\
  %Manual alignment (same task)              & \multicolumn{1}{c}{\textbf{1--4~hours}} \\
\end{tabular}
\end{ruledtabular}
\end{table}

Table~\ref{tab:timing} summarizes the timing of each alignment step. While the model inversion only takes a few milliseconds, the servo mirror motions take seconds. When using the Nelder--Mead refinement, an alignment needs 15--20 seconds due to multiple mirror motions. The spiral acquisition is needed only for a cold start far off the ridge and takes minutes. Note that the typical automatic alignment time of ${\sim}20$~s is still much faster than manual alignment. The qualitative difference is equally important: manual recovery can fail entirely when the retro signal is lost, resulting in a blind search in eight dimensions, whereas the model-based approach provides a model-based initial guess.
% ============================================================================
%  V. EXPERIMENTAL DEMONSTRATION
% ============================================================================
\section{\label{sec:experiment}Experimental Demonstration}

We demonstrate the framework experimentally using a cold-atom--cavity system. In particular (as shown in the schematic in Fig.~\ref{fig:optical_layout}), we steer a retro-reflected optical lattice at 784~nm that transports ${}^{87}$Rb atoms from a three-dimensional magneto-optical trap (3D MOT, $\pnt_2$) into the waist of a multimode optical cavity ($\pnt_1$). The conveyor belt is formed by ramping a 1~MHz frequency difference between the forward and retro-reflected beams, created by two crossed AOMs (IntraAction ATM-801A2). Further details are given in Appendix~\ref{app:layout}. Each of the four steering mirrors sits in an off-the-shelf Thorlabs Polaris kinematic mount retrofitted with low-cost, encoder-equipped serial-bus servos~(Feetech STS3032) that turn its two adjustment screws with an effective mirror-tilt resolution of ${\sim}3$--$10~\mu$rad (Appendix~\ref{app:hardware}).

Perhaps the most straightforward application is to move $\pnt_1$ around while keeping $\pnt_2$ fixed. This corresponds to transporting atoms from the fixed MOT position to different locations in the vacuum chamber, e.g. to find the position of strongest coupling to the cavity.

% ── Figure 5: Validation ──
\begin{figure}
  \includegraphics[width=\columnwidth]{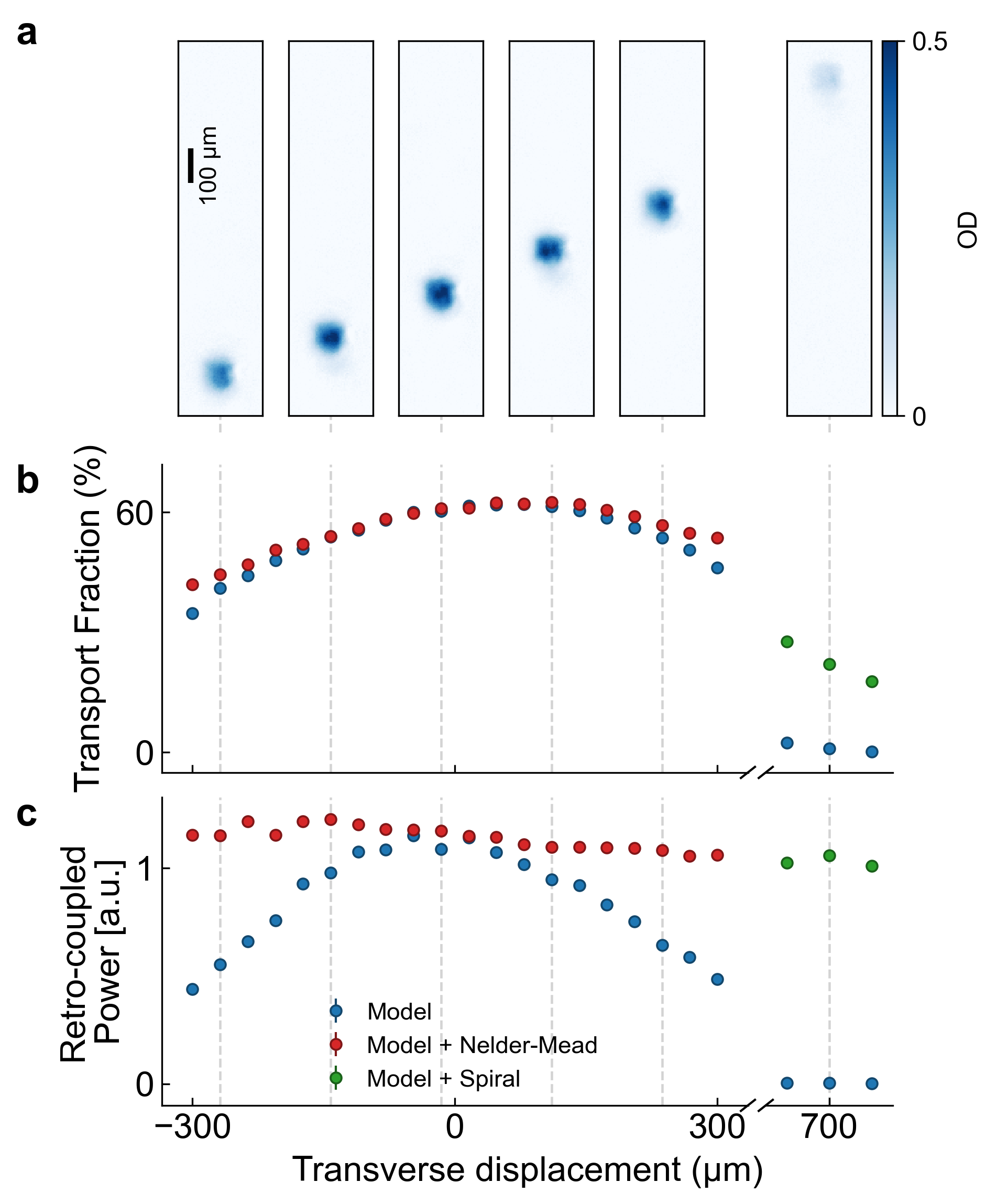}
  \caption{\label{fig:transport}\textbf{Model Validation.}
    Model-guided control holds the retro-reflected lattice aligned while the atom
    cloud is displaced transversely across the cavity plane after
    MOT-to-cavity transport.
    (a) Absorption images (optical density) show the
    cloud translating with the transverse displacement; dashed lines link each
    image to its position on the curves below.
    (b) The transport fraction (normalized to the initial MOT load) and
    (c) the retro-coupled power (a.u.) across the transverse-displacement scan; error bars (standard error of the mean over
    10 repetitions) are smaller than the markers.
    Blue: projective model alone; Red: optimizer layered on the model,
    improving the transport fraction and retro power most visibly at the
    largest transverse displacements.
    Green (right of the axis break, 650--750~$\mu$m, beyond the calibrated
    range): spiral-descent search (Sec.~\ref{subsec:optimize}) seeded by the
    model, where the model alone gives no retro signal.
  }
\end{figure}

Figure~\ref{fig:transport} shows such a transport experiment: The atom cloud's final transport position is displaced transversely across the cavity plane while keeping the transport distance constant. We confirm the cloud's displacement using absorption imaging [Fig.~\ref{fig:transport}(a)]. Note that the presence of a transported cloud at every displacement confirms that the retro-reflected beam remains co-aligned with the forward beam, the requirement for having a transport lattice at all, and that the lattice still intersects the MOT, from which the atoms are loaded. Over the central range the optical density is nearly unchanged, indicating that the lattice continues to pass through the center of the MOT; at large displacement (700~$\mu$m) the optical density is visibly reduced while the retro power is maintained, suggesting the lattice intersecting the MOT off-center and therefore loading fewer atoms. Quantitatively, we see that the transport fraction [Fig.~\ref{fig:transport}(b)] and the retro-coupled power [Fig.~\ref{fig:transport}(c)] stay within ${\sim}20\%$ of their peak values over the central $\pm150~\mu$m of the scan. Layering the Nelder--Mead optimizer on top of the model further improves the transport fraction and retro power, most visibly at the largest transverse displacements, where the model's accuracy decreases. Beyond the calibrated range, at 650--750~$\mu$m, the model alone produces limited retro signal or transport; the spiral-descent search seeded by the model recovers the retro power and a transport fraction of 18--28\%. The transport fraction at large displacements can be further improved by scanning the transverse displacement at the MOT plane or tuning the spacing parameter between the MOT and cavity planes.

% ============================================================================
%  VI. CONCLUSION AND OUTLOOK
% ============================================================================
\section{\label{sec:outlook}Conclusion and Outlook}

We have introduced a deterministic, model-based framework that solves optical alignment as an inverse problem: a compact projective-geometry forward model $g(\knob)$ maps the mirror angles to the beam coordinates at every target plane, and its numerical inverse $\knob^* = g^{-1}(\func^*)$ returns the mirror angles that satisfy all target constraints simultaneously. The model's role is not to predict the beam path to micron accuracy but to \textit{decorrelate the search space}---turning an ill-conditioned, coupled search into a well-conditioned search. Our method is suitable for otherwise prohibitively costly (in time or difficulty) alignment procedures.

We demonstrate the framework by aligning a four-mirror, retro-reflected lattice beam through an atom--cavity apparatus using low-cost motorized mounts and a single photodiode. Our model holds the retro-reflected lattice aligned while the atomic cloud position is scanned across the cavity plane. A single model inversion (${\sim}$ms) gives a near-optimal start and subsequently, a photodiode-fed refinement (15--20~s) reaches the optimum. Any optimized solution can later be recalled by interpolation.

By turning alignment from a manual bottleneck into a programmable, hands-off operation, the framework removes a longstanding barrier to scaling and automating experiments where optical access is limited or uptime is paramount---from optical lattice clocks and quantum-gas microscopes to fiber cavities and continuous atom--photon interfaces. The framework can also be adapted for fully automatic optical alignment, e.g. with AI-driven robotic platforms ~\cite{uddinAIDrivenRoboticsOptics2025}, and the long term goal of fully autonomously operating photonic labs.

All Python code (model, solver, hardware drivers), CAD files, firmware, and wiring schematics are available at \url{https://gitlab.com/simon-lab-public/motorizedmirror}.
% ============================================================================
%  ACKNOWLEDGMENTS
% ============================================================================
\begin{acknowledgments}
We thank Chi Shu for fruitful discussions and for the instrument design reported in his thesis~\cite{shu2022quantum}, which provided an important conceptual and technical foundation for the present work. 

This material is based upon work supported by the U.S. Department of Energy, Office of Science, National Quantum Information Science Research Centers. Support through Q-NEXT at SLAC National Accelerator Laboratory, under DOE Contract No.~DE-AC02-76SF00515 and Field Work Proposal No.~100645, enabled the development of the modeling approach presented here. We also acknowledge support from the U.S. Army Research Laboratory and the U.S. Army Research Office through the Multidisciplinary University Research Initiative (MURI) program under Award Nos.~W911NF-20-1-0136 and W911NF-25-1-0263, and from the U.S. National Science Foundation through the Challenge Institute for Quantum Computation (CIQC), via Agreement No.~00012167.

\end{acknowledgments}

\section*{Author Declarations}

\subsection*{Conflict of Interest}
J.S. acts as a consultant to and holds stock options from Atom Computing.

\subsection*{Author Contributions}
Bowen Li and Lukas Palm contributed equally to this work.

\textbf{Bowen Li:} Conceptualization (equal); Data curation (equal); Formal analysis (equal); Investigation (equal); Methodology (equal); Resources (equal); Software (equal); Supervision (equal); Validation (equal); Visualization (equal); Writing -- original draft (equal); Writing -- review \& editing (equal).
\textbf{Lukas Palm:} Conceptualization (equal); Data curation (equal); Formal analysis (equal); Investigation (equal); Methodology (equal); Resources (equal); Software (equal); Supervision (equal); Validation (equal); Visualization (equal); Writing -- original draft (equal); Writing -- review \& editing (equal).
\textbf{Xin Wei:} Conceptualization (equal); Investigation (equal); Validation (equal); Resources (equal); Methodology (equal); Software (equal); Supervision (equal); Writing -- review \& editing (equal).
\textbf{Marius J\"urgensen:} Resources (equal); Supervision (equal); Writing -- review \& editing (equal).
\textbf{Zeyang Li:} Resources (equal); Supervision (equal); Writing -- review \& editing (equal).
\textbf{Yiming Cady Feng:} Resources (equal); Writing -- review \& editing (equal).
\textbf{Abhishek V. Karve:} Resources (equal); Writing -- review \& editing (equal).
\textbf{Jon Simon:} Conceptualization (equal); Investigation (equal); Methodology (equal); Funding acquisition (lead); Project administration (lead); Resources (equal); Supervision (equal); Writing -- review \& editing (equal).

% ============================================================================
%  DATA AVAILABILITY
% ============================================================================
\section*{Data Availability Statement}
The data that support the findings of this study are available from the corresponding author upon reasonable request. All source code and hardware designs are openly available at
\url{https://gitlab.com/simon-lab-public/motorizedmirror}.

%============================================================================
%  APPENDICES
% ============================================================================
\appendix

% --- A: Translation and Rotation Matrices ---
\section{\label{app:matrices}Translation and Rotation Matrices}
 The $3\times3$ translation and rotation matrices used in the transform [Eq.~\eqref{eq:sandwich}] follow the conventions of Ref.~\onlinecite{corcovilosABCDsBetterMatrix2023}: $\Trans(u,v)$ translates a ray by $u$ in the $x$-direction and $v$ in the $y$-direction, and $\Rot(\varphi)$ rotates a ray counter-clockwise about the origin by $\varphi$,
\begin{equation}\label{eq:T}
  \Trans(u,v) = \begin{pmatrix}
    1 & -u & -v \\ 0 & 1 & 0 \\ 0 & 0 & 1
  \end{pmatrix}
  \quad
  \Trans^{-1}(u,v) = \Trans(-u,-v),
\end{equation}
\begin{equation}\label{eq:R}
  \Rot(\varphi) = \begin{pmatrix}
    1 & 0 & 0 \\
    0 & \cos\varphi & -\sin\varphi \\
    0 & \sin\varphi & \cos\varphi
  \end{pmatrix}
  \quad
  \Rot^{-1}(\varphi) = \Rot(-\varphi).
\end{equation} With these active transforms, the inner factors of Eq.~\eqref{eq:sandwich}, $\Rot^{-1}(\varphi)\,\Trans^{-1}(u,v)$, carry a lab-frame ray into the frame of an element located at $(u,v)$ and rotated by $\varphi$, where the canonical matrix $\RTM_0$ applies.

\section{\label{app:tangent}Treatment of the Tangential (Vertical) Plane}

\begin{figure}
  \includegraphics[width=\columnwidth]{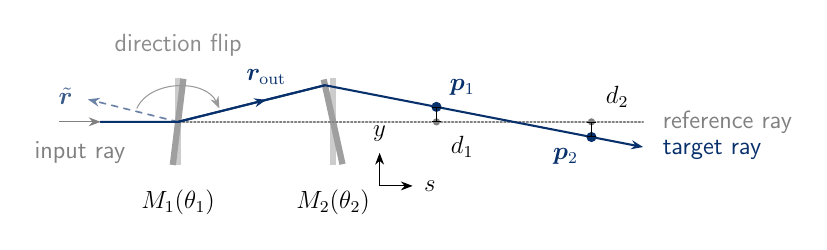}
  \caption{\label{fig:tangential}%
    \textbf{Two-mirror pair in the vertical (tangential) plane.} Side view of the pair of Fig.~\ref{fig:ellipse}(a) in unfolded path coordinates: $s$ runs along the beam and $y$ is the height, with the mirrors sitting on the reference axis (gray dotted). Faint plates mark the nominal orientation; solid plates are tilted by the vertical angles $\theta_1,\theta_2$. At $M_1$ the mirror matrix returns the reflected line $\tilde{\ray}$ (dashed), which runs back along the path; reversing the propagation direction maps it onto the forward line $\ray_\mathrm{out}$ of Eq.~\eqref{eq:y_mirror} through the same reflection point, so the beam (blue) continues with its vertical angle changed by $2\theta_1$. The same construction at $M_2$ adds $2\theta_2$, here of opposite sign, and the beam reaches the target points $\pnt_1,\pnt_2$ at the signed heights $d_1,d_2$ relative to the reference axis.}
\end{figure}

For the kinematic mirror mounts used here, the tangential plane is modeled in unfolded path coordinates: the path length $s$ along the beam replaces the table coordinate, so all elements lie on the reference axis and every fold mirror appears edge-on at normal incidence [Fig.~\ref{fig:tangential}]. A vertical-plane reflection is then implemented as the usual mirror transform followed by a reversal of the propagation direction. After applying the mirror matrix to obtain $\tilde{\ray}=(\tilde c,\tilde a,\tilde b)^T$, which runs back along the path, we intersect this ray with the mirror line to find $\pnt_m=(1,x_m,y_m)^T$ and rewrite the outgoing line as
\begin{equation}\label{eq:y_mirror}
  \ray_\mathrm{out}
  =
  \begin{pmatrix}
    -\left(y_m + m x_m\right) \\
    m \\
    1
  \end{pmatrix},
  \qquad
  m=-\frac{\tilde a}{\tilde b}.
\end{equation}
This is $\tilde{\ray}$ reflected about the vertical through $\pnt_m$: the slope $m$ of the back-traveling line becomes $-m$, so the beam's vertical angle is carried through the fold and changed only by the tilt-induced deflection $2\theta_i$ [Fig.~\ref{fig:tangential}]. Since homogeneous ray vectors are defined only up to scale, the overall sign of $\ray_\mathrm{out}$ is immaterial, and we normalize to $b=1$.

% ── Figure: tangential-plane two-mirror pair (Appendix B) ──

Additionally, since the motorized adjusters are not arranged as a true gimbal, a tilt in one axis also translates the effective reflection point seen by the orthogonal plane. In the model this is treated as a geometric compensation: for a mirror with adjuster separation $d_m$ and already-solved orthogonal angle $\theta_\perp$, the reflection point is displaced by $\Delta r = (d_m/2)\tan(-\theta_\perp)$ along the corresponding mirror-mount axis before the ray matrices are evaluated. The solver therefore performs a short fixed-point loop: solve one plane, pass the resulting mirror angles as reflection-point offsets to the orthogonal-plane solve, and then re-solve the first plane with the reciprocal offsets included. Because the coupling enters only through the small reflection-point shift, a single reciprocal pass suffices at small tilt angles; at large angles the initial mismatch $\Delta r$ grows with $\tan\theta_\perp$ and the loop is iterated until the mirror angles stop changing. Any residual is absorbed by the downstream photodiode optimization.

% --- Jacobian structure ---
\section{\label{app:jacobian}Jacobian Structure of a Two-Mirror Pair}

\begin{figure}
  \includegraphics[width=\columnwidth]{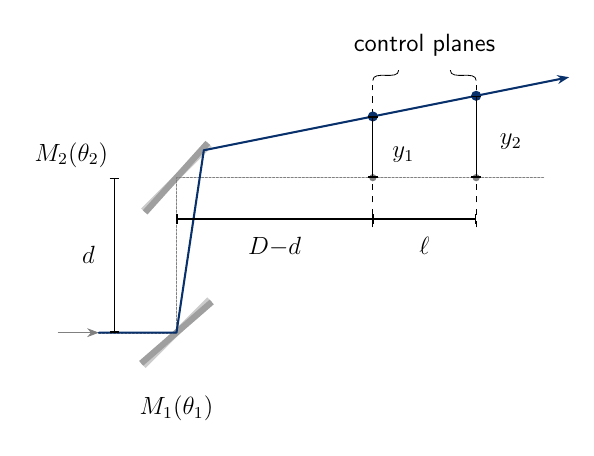}
  \caption{\label{fig:jacobian}%
    \textbf{Jacobian structure of a two-mirror pair.}
    Tilting the mirrors by $\theta_1$, $\theta_2$ (solid plates; nominal mirrors faint) walks the beam (blue) off the reference path (dotted), producing the beam-space offsets $y_1$, $y_2$ of Eq.~\eqref{eq:jac} at the two control planes (dashed). Each tilt $\theta_i$ deflects the beam by $2\theta_i$, which the remaining path length converts into an offset: the lever arms $D$, $D{-}d$, $D{+}\ell$, and $D{+}\ell{-}d$ of the four Jacobian entries differ only by the small separations $d$ and $\ell$, giving the near-singular structure of Eqs.~\eqref{eq:detjac} and \eqref{eq:kappa}.
  }
\end{figure}

For a two-mirror pair with inter-mirror path distance $d$, let $D$ be the path distance from $M_1$ to the first control plane and $\ell$ the separation between the two control planes. Unfolding the reflections, a tilt $\theta_i$ of mirror $M_i$ leaves the beam position at that mirror unchanged and injects a pure angle $2\theta_i$---exact for in-plane tilts---which the remaining path converts into a transverse offset through its lever arm. Accumulating both kicks at each plane (Fig.~\ref{fig:jacobian}) gives the beam-space offsets of Sec.~\ref{sec:knobfunc},
\begin{align*}
  y_1 &= 2D\,\theta_1 + 2(D-d)\,\theta_2 \\
  y_2 &= 2(D+\ell)\,\theta_1 + 2(D+\ell-d)\,\theta_2
\end{align*}
so that the Jacobian is
\begin{equation}\label{eq:jac}
  \Jac = \frac{\partial\func}{\partial\knob}
    = 2\begin{pmatrix}
        D & D-d \\ D+\ell & D+\ell-d
      \end{pmatrix}.
\end{equation}
Both mirrors act through nearly the same lever arm---the two columns differ only by the small separation $d$---and the determinant collapses to
\begin{equation}\label{eq:detjac}
  \det\Jac = 4\,d\,\ell,
\end{equation}
independent of the common distance $D$.  With $\sigma_\mathrm{max}\approx 4D$ for $d,\ell \ll D$, and $\sigma_\mathrm{min} = |\det\Jac|/\sigma_\mathrm{max}$, the condition number is
\begin{equation}\label{eq:kappa}
  \kappa = \frac{\sigma_\mathrm{max}}{\sigma_\mathrm{min}}
         = \frac{\sigma_\mathrm{max}^2}{|\det\Jac|}
         \approx \frac{4D^2}{d\,\ell},
\end{equation}
which reduces to $\kappa\sim D/d$ only when the control planes are separated by $\ell\sim D$; for closely spaced control points the ill-conditioning is correspondingly worse. For cases where $\frac{4D^2}{d\,\ell}\gg 1$ this produces the elongated response ellipse of Fig.~\ref{fig:ellipse}(b), whose aspect ratio is $\kappa$; the model-based inversion effectively applies $\Jac^{-1}$, mapping the ellipse to a circle.

\section{Measuring the Ridge Direction}
\label{app:spiral}
The Nelder--Mead refinement of Sec.~\ref{subsec:optimize} runs in the ridge eigen-frame of a mirror pair, which is either predicted by the model or measured. The direct measurement is a two-dimensional scan of the pair's two knobs: sample $P(\knob)$ on a grid, keep the points above half maximum, and form the intensity-weighted mean and covariance
\begin{align}\label{eq:eigenframe}
  \bar{\knob} &= \frac{\sum_k w_k\,\knob_k}{\sum_k w_k},
  \quad
  \mat{C} = \frac{\sum_k w_k\,(\knob_k-\bar{\knob})(\knob_k-\bar{\knob})^{\!\top}}
                 {\sum_k w_k}\\
  w_k &= P_k - P_\mathrm{min},
\end{align}
with $P_\mathrm{min}$ the smallest retained power. The eigenvectors of $\mat{C}$ estimate the along- and across-ridge directions $\hat{\vect{e}}_\parallel$, $\hat{\vect{e}}_\perp$, and the square root of the eigenvalue ratio estimates the ridge anisotropy. Such a scan resolves the ridge, but it spends most of its $N\times N$ points on the dark background, where $w_k\approx0$, and on long motor moves.

The spiral search of Algorithm~\ref{alg:spiral} typically estimates the ridge direction with fewer measurements than a full two-dimensional grid scan. Its adaptive recentering directs sampling toward higher-power regions, reducing measurements spent on the dark background. In our implementation, a few tens of samples collected over a few turns are typically sufficient for this initialization. We then start the Nelder--Mead simplex from the best sampled setting, with its long axis aligned along $\hat{\vect{e}}_\parallel$ and its aspect ratio set by the estimated anisotropy.
% --- Hardware ---
\section{\label{app:hardware}Motorized Kinematic Mounts}

\begin{figure*}[t]
  \includegraphics[width=\textwidth]{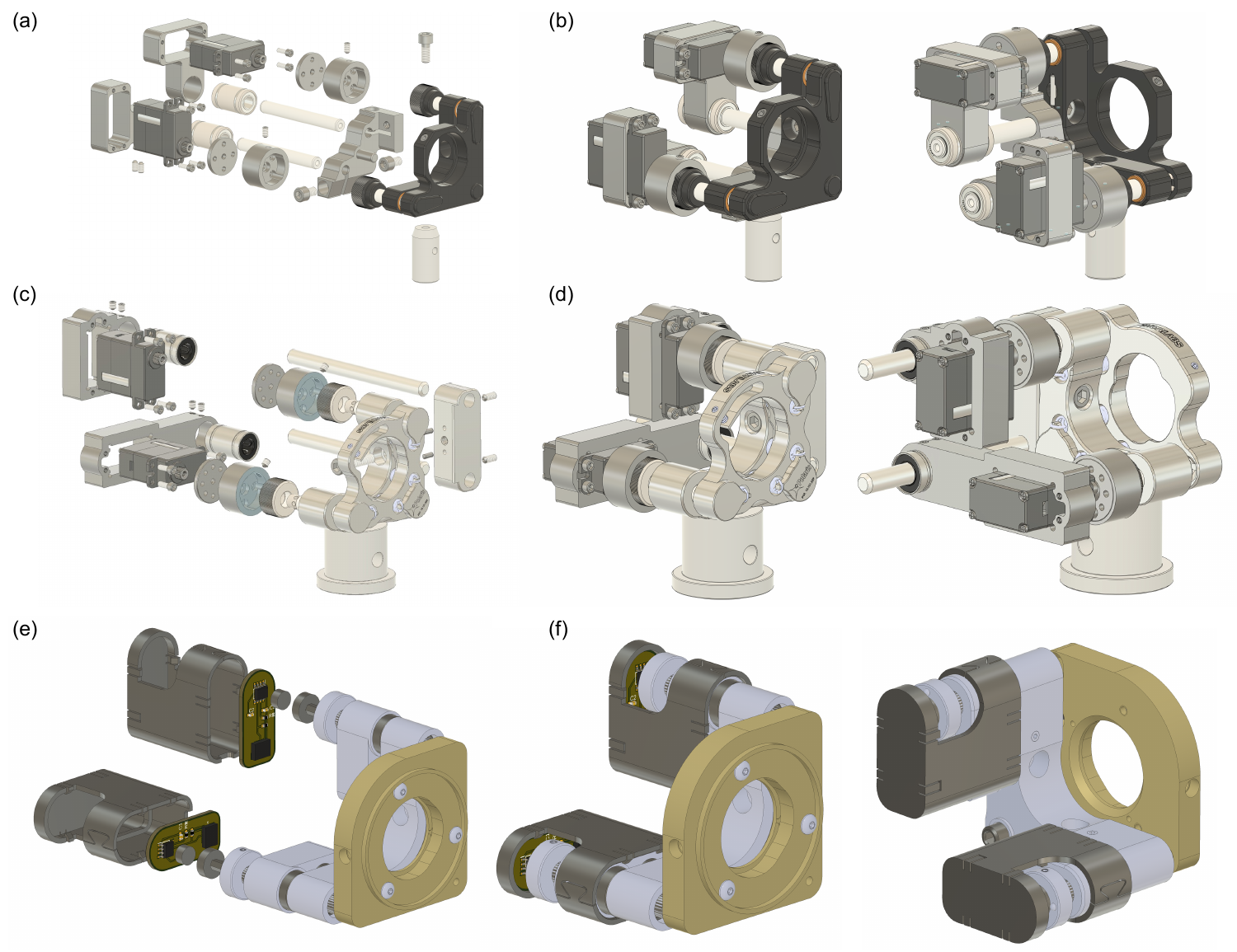}
  \caption{\label{fig:motor_cad}%
    Drawings of motorized kinematic mounts. \textbf{(a)--(b)} Thorlabs KM100 mounts. \textbf{(c)--(d)} Polaris mounts. \textbf{(e)--(f)} Piezo-motor mount with retrofit encoders.
  }
\end{figure*}

Two mount types are used (Fig.~\ref{fig:motor_cad}):

\textbf{Servo Polaris mount:}
Thorlabs Polaris kinematic mount with a CNC coupling adapter and STS3032 serial-bus servo. A linear bearing is used on the Thorlabs cage rod to constrain the motion of the thread adjuster. Eight channels are addressed on a single UART daisy chain. The servo's built-in absolute encoder provides position feedback with 12-bit angular resolution (4096 counts/$360^\circ$); the fine-pitch adjustment screw gives a step of $1.8~\mu$rad of mirror tilt per count and a unidirectional repeatability of ${\sim}3$--$10~\mu$rad~\cite{shu2022quantum}, which we quote as the effective mirror-tilt resolution. Cost: ${\sim}\$200$/axis. Passive hold via gear friction. One-turn travel time: ${\sim}2$~s. Backlash effects are reduced by approaching each target from the same direction.

\textbf{Piezo-motor mount with retrofit encoders:}
Picomotor-actuated mirror mounts (Newport New Focus 8816-6) with absolute rotary encoders (AS5600) retrofitted to the adjustment screws for position feedback. We also provide the PCB and 3D-printed/CNC mount adapter designs for this custom solution. Better encoders (e.g., AS5048A, 14-bit) would further improve resolution. Mirror-tilt resolution: ${\sim}2~\mu$rad. Cost: ${\sim}\$300$/axis. Passive hold via the friction of the adjustment screw (the Picomotor holds position unpowered). One-turn travel time: 15~s.

\begin{table}[h]
\caption{\label{tab:mounts}Mount comparison.}
\begin{ruledtabular}
\begin{tabular}{lcc}
  Property & Servo Polaris & Piezo + encoder \\
  \colrule
  Mirror-tilt resolution  & ${\sim}3$--10~$\mu$rad & ${\sim}2~\mu$rad \\
  One turn travel time   & ${\sim}2$~s       & 15~s          \\
  Cost/axis            & ${\sim}\$200$          & ${\sim}\$300$     \\
  Position feedback    & built-in (servo)     & retrofit encoder \\
  Passive hold         & yes (gear)           & yes (friction)   \\
\end{tabular}
\end{ruledtabular}
\end{table}

% ── Figure Motor ──

% --- Layered Control Architecture ---
\section{\label{app:architecture}Layered Control Architecture}

\begin{figure*}
  \includegraphics[width=0.8\textwidth]{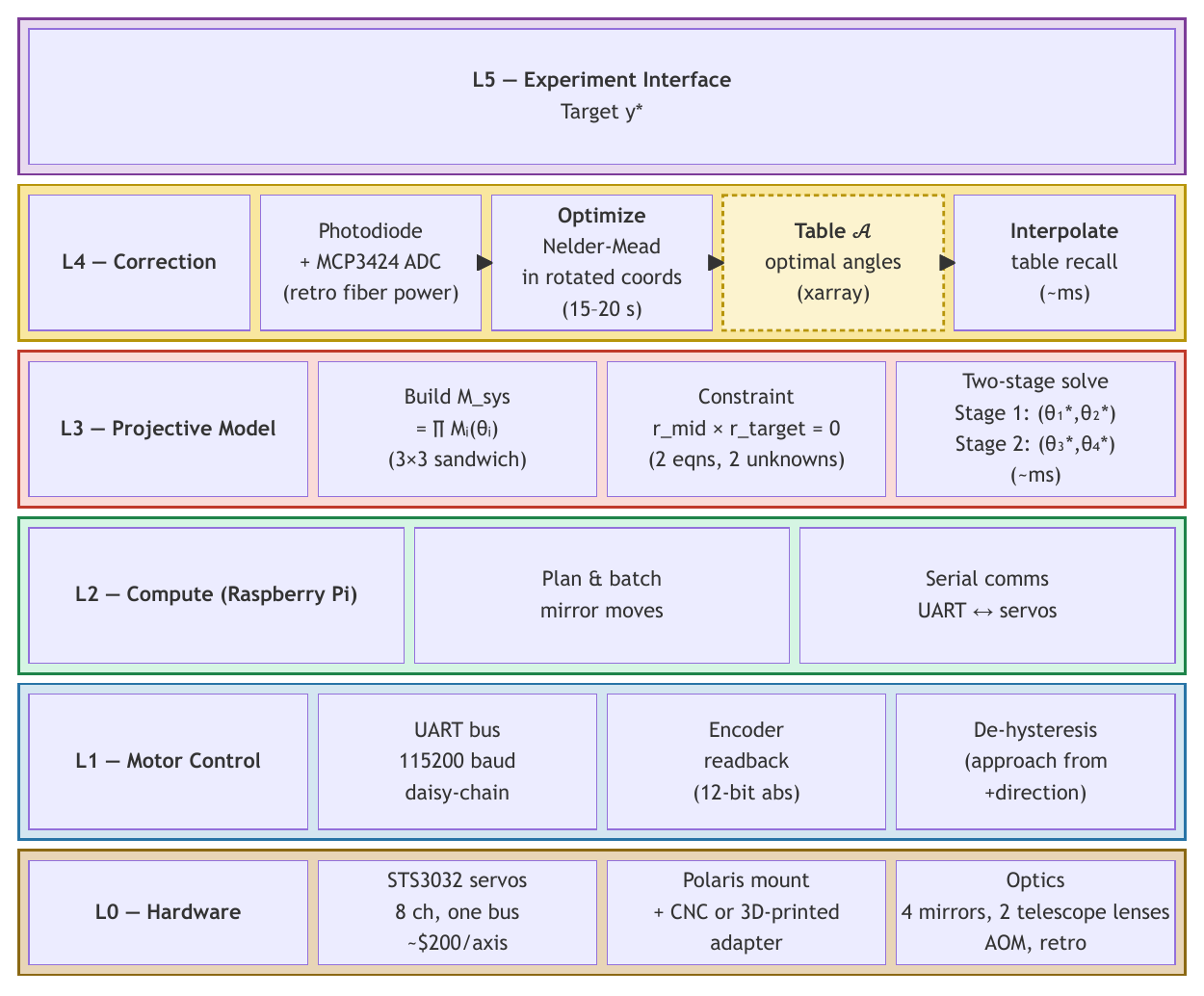}
  \caption{\label{fig:layer_diagram}%
    Layered control architecture (L0--L5).
    \textbf{L0 (Hardware):} STS3032 servos on Polaris mounts with CNC or 3D-printed
    adapters drive four mirrors, or Picomotors with encoders.
    \textbf{L1 (Motor Control):} UART daisy chain at 115\,200 baud with
    12-bit absolute encoder readback and uni-directional approach for backlash
    elimination.
    \textbf{L2 (Compute):} Raspberry Pi batches and plans mirror moves,
    and handles serial communication.
    \textbf{L3 (Projective Model):} Builds the system matrix
    $\mathbf{M}_{\mathrm{sys}}=\prod_i M_i(\theta_i)$ via the $3\times3$
    sandwich transform, solves the collinearity constraint
    $\mathbf{r}_{\mathrm{mid}}\times\mathbf{r}_{\mathrm{target}}=0$ in two
    sequential stages (${\sim}$ms).
    \textbf{L4 (Correction):} the Optimize step runs a Nelder--Mead optimization
    in rotated mirror-pair coordinates against the retro-coupled fiber power from
    an MCP3424 ADC (15--20\,s) and populates a pre-measured \texttt{xarray}
    table of optimal angles; the Interpolate step serves a new target by
    multilinear interpolation of that table (${\sim}$ms, no optimization).
    \textbf{L5 (Experiment Interface):} Accepts target positions
    $\mathbf{y}^{*}$.
  }
\end{figure*}

The full software/hardware stack that implements the deterministic solver and the measurement-based correction is summarized in
Fig.~\ref{fig:layer_diagram}, organized into six layers (L0--L5).

% ── Layered Architecture ──

% --- Optical layout plots ---
\section{\label{app:layout}Optical Layout Plots}

\begin{figure*}[t]
  \includegraphics[width=\textwidth]{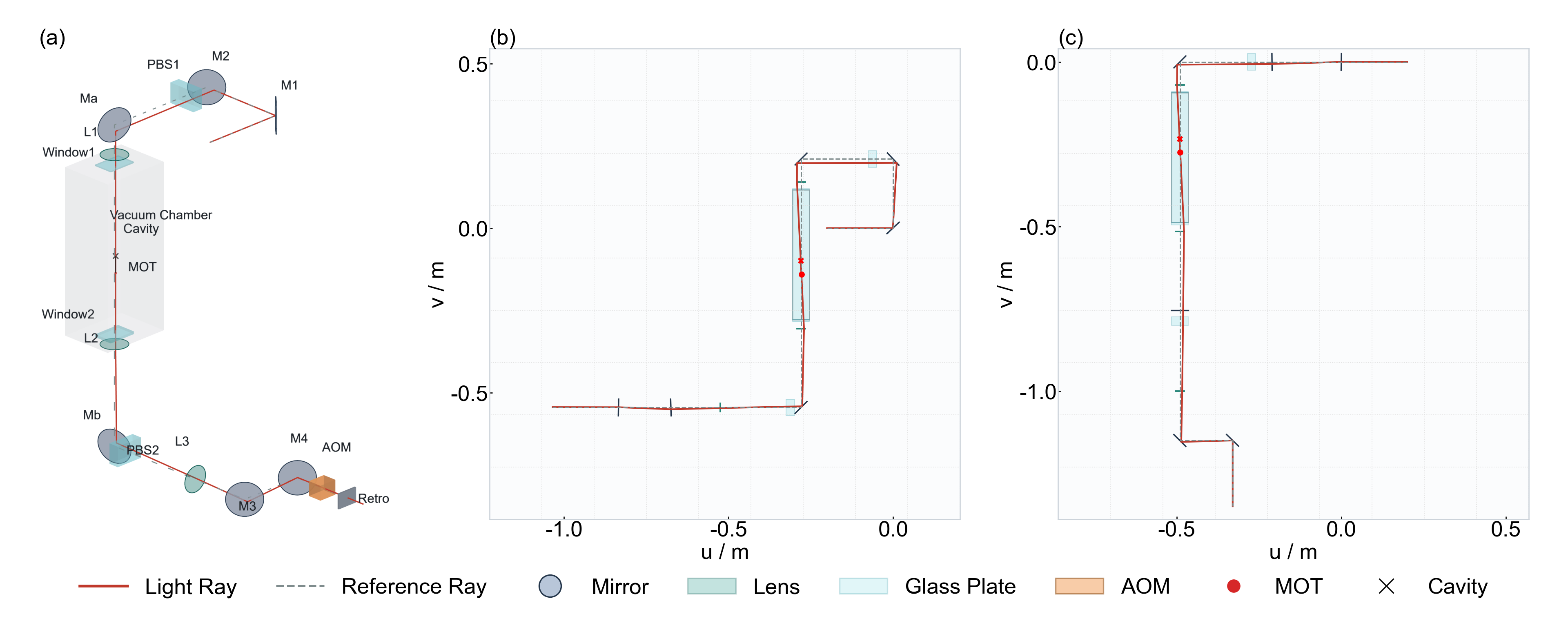}
  \caption{\label{fig:optical_layout}%
    Schematic of the optical layout with motorized kinematic mounts. \textbf{(a)} Three-dimensional layout of the folded beam path through the four actuated mirrors (M1--M4), turning mirrors (Ma, Mb), telescope lenses (L1--L3), vacuum-chamber windows, and the vacuum chamber, with the solved beam (red) and the nominal reference beam (gray dashed) meeting the MOT and cavity targets. \textbf{(b)},~\textbf{(c)} Projections of the same solution onto the two orthogonal solver planes.
  }
\end{figure*}

Figure~\ref{fig:optical_layout} shows the full solved beam path for the apparatus of Sec.~\ref{sec:experiment}, together with its projections onto the two orthogonal solver planes.

% ============================================================================
%  BIBLIOGRAPHY
% ============================================================================
\clearpage  % flush the remaining appendix figures so the reference list comes last
\renewcommand{\refname}{References}  % the AIP reprint style suppresses this heading; restore it
\bibliography{references}

\end{document}